\documentclass[11pt,notitlepage]{isasrep}
\usepackage[dvips]{graphicx}
\usepackage[small,hang]{caption2}
\def\apj{{ApJ}}

\def\mnras{{MNRAS}}

\def\aap{{A\&A}}

\begin{document} 

%%% Title of your paper %%%	                                           

\title{Galaxy populations from Deep \\ 
  \iso\ Surveys\footnote{To appear in ``Mid- and Far-infrared
  Astronomy and Future Space Missions'',  ISAS Report Special
  Edition (SP No. 14), ed. T. Matsumoto \& H. Shibai }}
				
%%% Running title of your paper %%%

\markright{A Template of the Manuscripts}

%%% Authors %%%

\author{
     Seb {\sc Oliver}\footnotemark[1] %,$\;$
%     Second {\sc Author}\footnotemark[1]$\;$
%     and		
%     Third M. {\sc Author}\footnotemark[2] 
} 

%%% Date of Submission %%%

\date{(11 August 2000)} 

\maketitle	%% DO NOT CHANGE this line %%

%%% INPUT your affilations as the footnotes %%%

\renewcommand{\thefootnote}{\fnsymbol{footnote}} %% DO NOT CHANGE this line %%

\footnotetext[1]{University of Sussex, Brighton, BN1 9QJ, UK.}
%\footnotetext[2]{3rd Author's address 999--9999, COUNTRY.}

\renewcommand{\thefootnote}{\arabic{footnote}}   %% DO NOT CHANGE this line %%
\setcounter{footnote}{0}			 %% DO NOT CHANGE this line %%
\renewcommand{\baselinestretch}{1}		 %% DO NOT CHANGE this line %%

%%% INPUT abstract here %%%

\begin{abstract}
I discuss some of the main extra-galactic field surveys which have been
undertaken by the Infrared Space Observatory (\iso).  I review the
findings from the source counts analysies and then examine some 
of the more recent detailed investigations into the explicit 
nature of the populations that make up these source counts.  

\end{abstract}

%%% Main body of the paper %%%

\section{\iso\ Surveys}
\begin{figure}[ht] 
   \begin{center}
   \includegraphics[angle=0,width=11cm,clip]{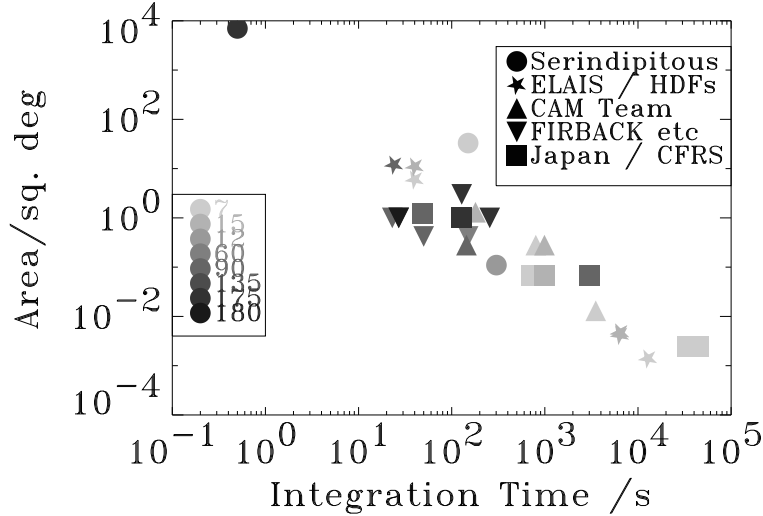}
   \caption{Comparison of \iso\ surveys: area vs. depth.  All the
surveys in Table \ref{tab:surveys} are plotted.  
 (Adapted from \protect\cite{Oliver et al. 2000})}
   \label{isosurveys}
   \end{center}
\end{figure}

Depending on how you count them, \iso\ (\cite{Kessler et al. 1996})
undertook around fifteen
independent field surveys to explore the extra-galactic populations.
These surveys primarily used the \iso-PHOT (\cite{Lemke et al. 1996})
and \iso-CAM (\cite{Cesarsky et al. 1996}) instruments.
As shown in Table \ref{tab:surveys} and illustrated in Figure \ref{isosurveys} these  surveys explored the 
full extent of \iso's parameter space in terms of wavelength, area and depth.  

\begin{table}
\begin{center}
\caption{Field Surveys with {\protect\em ISO}, ordered roughly in decreasing
area. References: 
1 -- \cite{Bogun et al. 1996}, 
2 --   \cite{Siebenmorgen et al. 1996}, 
3 -- \cite{Oliver et al. 2000} 
4 -- \cite{Elbaz et al. 1999}, 
5 --  \cite{Dole et al. 1999},
6 -- \cite{Juvela et al. 2000},
7 -- \cite{Linden-vornle 2000}, 
8 -- \cite{Elbaz et al. 1999}
9 --  \cite{Clements et al. 1996},
10 --  \cite{Flores et al. 1999a},\cite{Flores et al. 1999b}, 
11 --  \cite{Elbaz et al. 1999}, 
12 -- Oliver et al. (2000, in preparation), 
%13 --  \cite{Taniguchi et al. 1997a}, 
13 --  \cite{Taniguchi et al. 1997}, 
14 --  \cite{Kawara et al. 1998}, 
15 --  \cite{Serjeant et al. 1997}}\label{tab:surveys}

\begin{tabular}{lllll}\hline \hline
%\hline
Survey Name  & [e.g. ref] & Wavelength & Integration & Area\\
             &    &   $/\mu$m  &   $/$s      & $/{\rm sq deg}$\\
\hline
PHT Serendipity Survey 	& 1  & 175         & 0.5            & 7000 \\
CAM Parallel Mode 	& 2    & 6.7           & 150            & 33 \\
ELAIS          	        & 3& 6.7,15,90,175 & 40, 40, 24, 128& 6, 11, 12,1\\
CAM Shallow   		& 4      & 15          & 180            & 1.3  \\
FIRBACK      		& 5& 175         & 256, 128       & 1, 3  \\
CIBR        		& 6      & 90, 135,180 & 23, 27, 27     & 1, 1, 1 \\
SA 57         		& 7& 60, 90      & 150, 50        & 0.42,0.42  \\
CAM Deep      		& 8 & 6.7, 15, 90   & 800, 990, 144  & 0.28, 0.28, 0.28\\
Comet fields       	& 9& 12          & 302            & 0.11\\
CFRS          		& 10                   & 6.7,15,60,90  & 720, 1000, 3000,3000 & 0.067.0.067.0.067,0.067\\
CAM Ultra-Deep		& 11      & 6.7           & 3520           & 0.013 \\
ISOHDF South  		& 12         & 6.7, 15       &$>6400, >6400$  & 4.7e-3, 4.7e-3 \\
%Deep SSA13              & 13               & 6.7           & 34000          & 2.5e-3\\
Deep Lockman            & 13,14      & 6.7, 90, 175  & 44640, 48, 128 & 2.5e-3, 1.2 , 1 \\
ISOHDF North            & 15       & 6.7, 15       & 12800, 6400    & 1.4e-3, 4.2e-3 \\
\hline
\end{tabular}
\end{center}
\end{table}

This is an attempt to collect together some of the findings of
these surveys.  I will group the surveys by wavelength,
since the populations explored at different wavelengths are
expected to be somewhat different.

\subsection{Mid IR surveys 6.7 - 15$\mu$m}
Surveys in this wavelength regime tend to pick up emission from 
hotter sources such as AGN or luminous starbursts, they also
pick up ``cirrus'' emission from normal galaxies.  While
these bands are located far from the peak of the typical starbursts
spectral energy distribution it has been noted that the PAH
features which do fall in these bands can be used as a good measure of
star-formation activity.
Surveys in these bands are the HDF surveys in the North and South;
CFRS; ELAIS; the ISOCAM guaranteed time Extra Galactic Surveys, the  Comet 
fields, the Japanese Lockman and Selected Area surveys and the CAM parallel 
mode survey.

\subsection{Far IR surveys 50-100$\mu$m}
These wavelengths are located at the peak of the rest-frame emission from
obscured star formation and thus provide the most direct estimate of the 
obscured bolometric output in star-forming galaxies.  Surveys in this
band are that of SA57; the Japanese surveys; ELAIS; CIBR and CFRS.

\subsection{``Cool'' FIR surveys 120-200$\mu$m}
I use this term to denote bands beyond the normal peak in a star-burst
emission.  Surveys at these bands tend to pick up galaxies which are
``cool'' either because the energy distributions have been ``cooled'' simply
by virtue of their high redshift, or because they are intrinsically cool
but local.
Surveys in this band are FIRBack; ELAIS, Japanese surveys; the PHOT slew
surveys; and CIBR.

\section{Counts}
\subsection{Mid Infrared}

\begin{figure}[ht] 
\vspace{-3cm}
   \begin{center}
\hspace{-1.5cm}
   \includegraphics[angle=0,width=8cm]{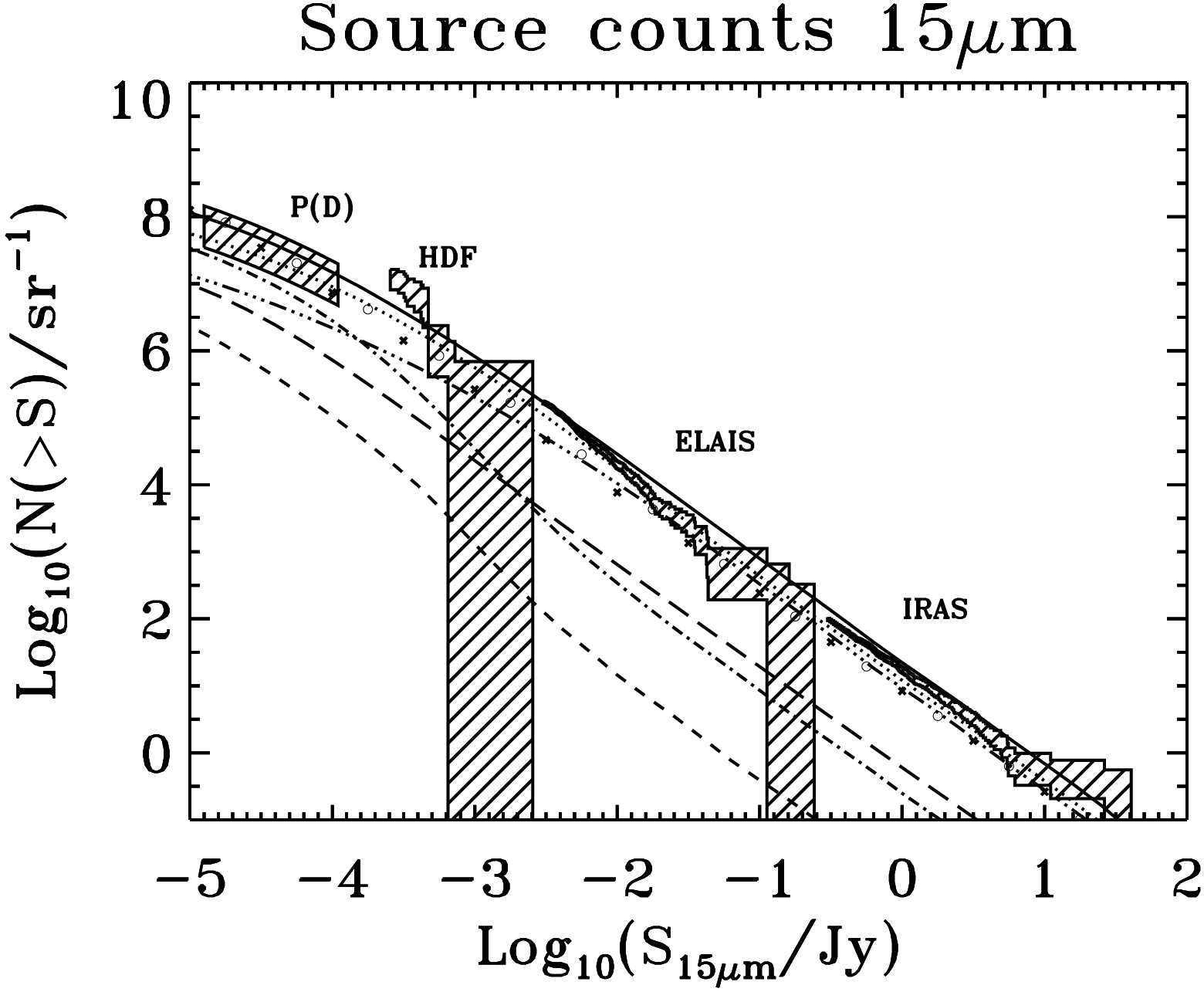}
\hspace{-5.5cm}
\includegraphics[angle=0,width=14cm,clip]{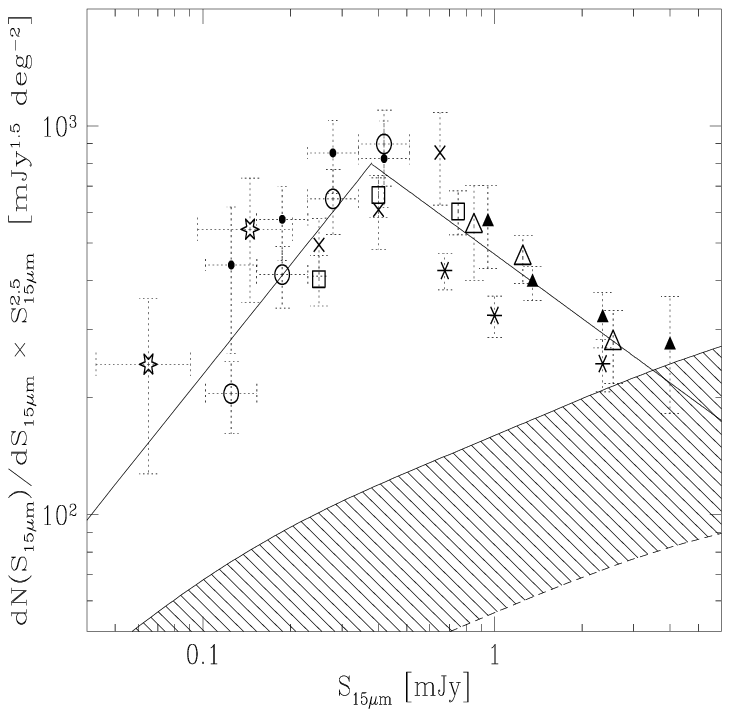}

   \caption{Number counts of 15$\mu$m galaxies.
Left: integral counts. Shaded regions show the ranges spanned by $\pm1\sigma$ uncertainties.
Also shown are the source counts and P(D) analysis from the Hubble
Deep Field \cite{Oliver et al. 1997}, \cite{Aussel et al. 1999}.
The \iras\ counts are estimated from
the $12\mu$m counts.
The  \cite{Franceschini et al. 1994} source count model
is over-plotted.  
The model spiral contribution is shown as a dotted line, ellipticals
as a dashed line, S0 as a dash-dot line, star-bursts as a
dash-dot-dot-dot line and AGN as a long dashed line. 
The total population model is shown as a full line. 
Also plotted are the \protect\cite{Guiderdoni et al. 1998} models A and E, as
small filled crosses and small open circles respectively. 
Figure taken from \protect\cite{Serjeant et al. 2000}
Right:
differential, Euclidean normalised counts. Data points: A2390 (open stars), \iso--HDF North (open circles), \iso--HDF-South (filled circles), Marano FIRBack (MFB) Ultra-Deep (open squares), Marano Ultra-Deep (crosses), MFB Deep (stars), Lockman Deep (open triangles), Lockman Shallow (filled triangles).  The hatched area shows the range of expectations
assuming no evolution.  Figure Taken from \protect\cite{Elbaz et al. 1999}}
   \label{fig:15counts}
   \end{center}
\end{figure}

The ELAIS 15$\mu$m counts \cite{Serjeant et al. 2000}, illustrated in Figure 
\ref{fig:15counts},  were shown to agree with many strongly evolving 
models such as those of \cite{Franceschini et al. 1994},
\cite{Pearson and Rowan-Robinson 1996}, \cite{Xu et al. 1998}
which had been based on \iras\ counts,  while being inconsistent with at least one
non-evolving model that of \cite{Xu et al. 1998}. Deeper counts such as those of the
\iso--HDF \cite{Oliver et al. 1997} amplify this picture of strongly evolving populations and the summary presented by \cite{Elbaz et al. 1999} (also shown in Figure \ref{fig:15counts}) suggests that the deepest cluster--lens surveys have seen a turn over in the counts and thus the beginning of the end of these evolving populations.

\subsection{FIR Counts}
\begin{figure}
\centering
\includegraphics[width=.6\textwidth]{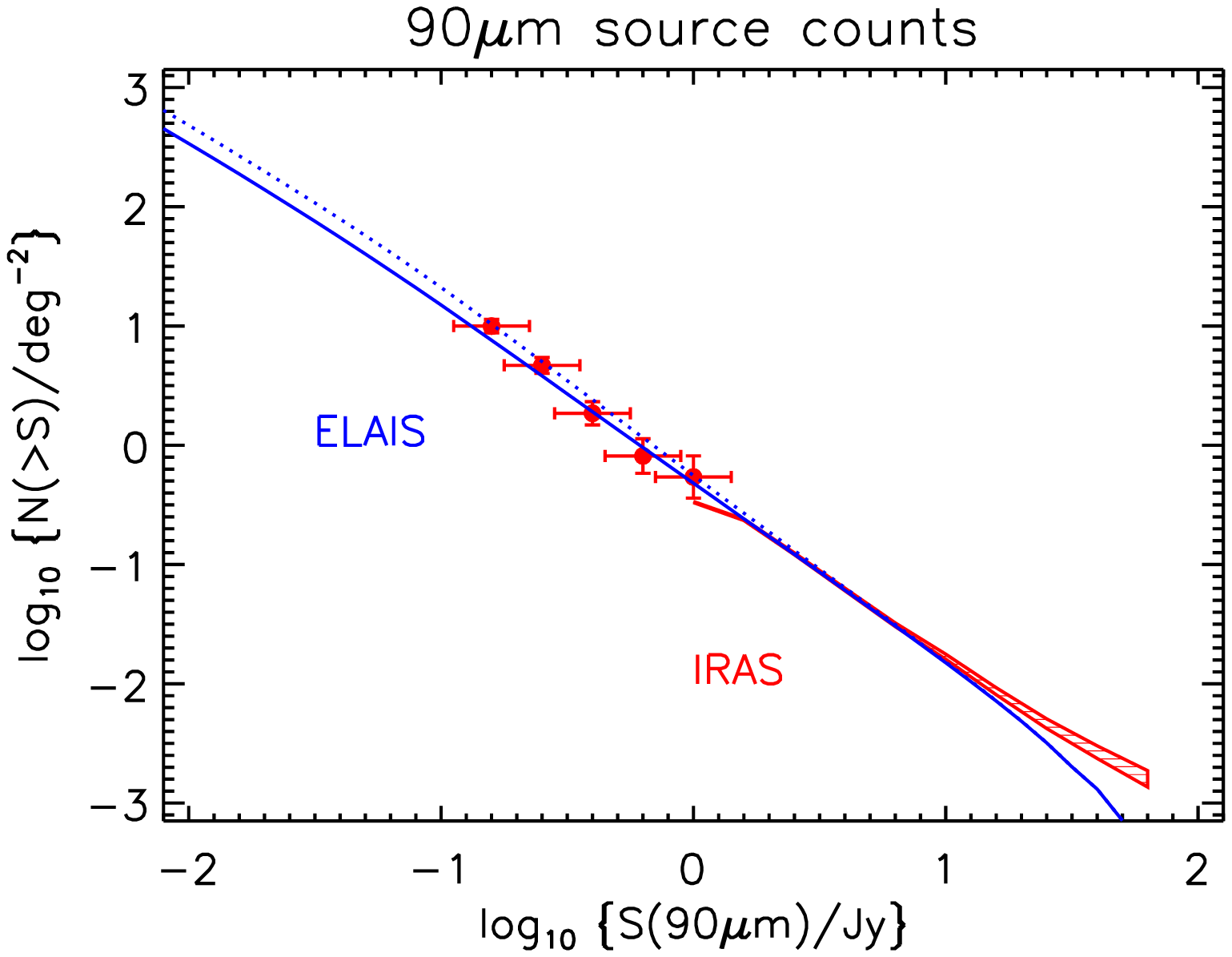}
\caption[]{
ELAIS and \iras\ 90$\mu$m source counts.  The solid line
and dotted lines are the counts predicted by the models A and E (respectively)
of \protect\cite{Guiderdoni et al. 1998}
Taken from \protect\cite{Efstathiou et al. 2000} 
}\label{fig:90counts} 
\end{figure}

The FIR source counts also show good evidence for evolution.  E.g. the ELAIS
90$\mu$m counts of \cite{Efstathiou et al. 2000} confirm the strong evolution expected from extrapolation from \iras\ studies, see Figure \ref{fig:90counts}.  
FIR counts from other \iso\ surveys are given in \cite{Juvela et al. 2000} and
\cite{Linden-vornle 2000}.

\subsection{``cool'' FIR counts}
\begin{figure}
%\begin{minipage}{5.5cm}
\centering
\includegraphics[width=0.8\textwidth]{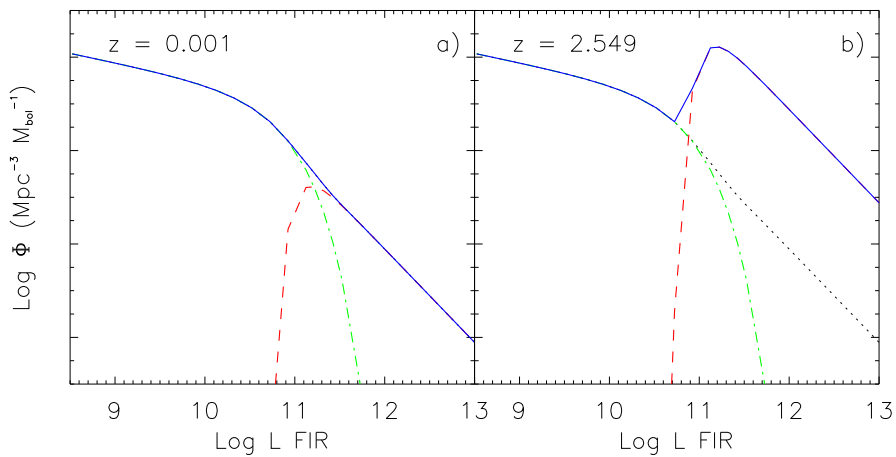}
\vspace{-2.0cm}
\caption[]{Two snapshots of a model fitting the 175$\mu$m counts and the FIR background, taken from \cite{Dole et al. 2000}. {\em a:} Luminosity Function at $z=0$ (solid line); normal galaxy (dot-dash); ULIRG
(dash-dash). {\em b}: Luminosity Function at $z=2.5$ (solid line); normal galaxy (dot-dash) ULIRG
(dash-dash) and local LF (dots). }
\label{lf_z0}

\end{figure}

The 175$\mu$m counts have attracted a lot of attention since they are more difficult to explain e.g. \cite{Dole et al. 1999} and references therein.  The counts exceed models
extrapolated from \iras.  To illustrate this Figure \ref{lf_z0} shows the \cite{Dole et al. 2000} attempt to explain the ``cool'' counts. This model allowed a strong evolution in the ultra luminous \iras\ galaxies and was constrained by both the 175$\mu$m counts and the FIR background.  It should be stressed that such an extreme model is not a unique fit to the data, but illustrates the lengths to which one needs to go to explain these data.

\section{Beyond Counts}
The counts are challenging, they show strong evolution, and may suggest new populations.  However, in isolation they do not tell us what the nature of the sources is.  The current effort is now strongly focused on more detailed investigations of these populations.  I pick three particular results for discussion.

\begin{table}
\centering
\caption[]{Provisional spectral classifications of \iso\ and 21cm
selected sources in the Southern ELAIS field S1.  Spectra come
from a one hour 2dF exposure and a number of nights on the ESO 3.6m and
NTT telescopes (Gruppioni et al., in prep. and La Franca et al., in prep.). Table taken from \protect\cite{Oliver et al. 2000a}}\label{tab:class}
\begin{tabular}{lrrrr}\hline\hline
Class & \multicolumn{2}{c}{21cm}& \multicolumn{2}{c}{\em ISO} \\
      &   2dF        &    ESO      &   2dF    & \multicolumn{1}{c}{ESO} \\\hline
Absorption 	& 48    &   	&15	& 6	\\
Star-burst 	& 9	&3	&22	&52 \\
H$\alpha$	&5	&2	&26	&\\
OII		&10	&	&4	&\\
OIII		&1	&	&	&\\
AGN/QSO		&6	&2	&20	&19\\
AGN/Sy1		&2	&1	&3	&\\
AGN/Sy2		&8	&3	&8	&8\\
AGN/BLLac	&	&	&2	&\\
Stars		&1	&	&8	&3\\
Too Faint	&60	&3	&41	&\\\hline
\end{tabular}
\end{table}

\begin{figure}
\begin{center}
{\includegraphics[width=10cm]{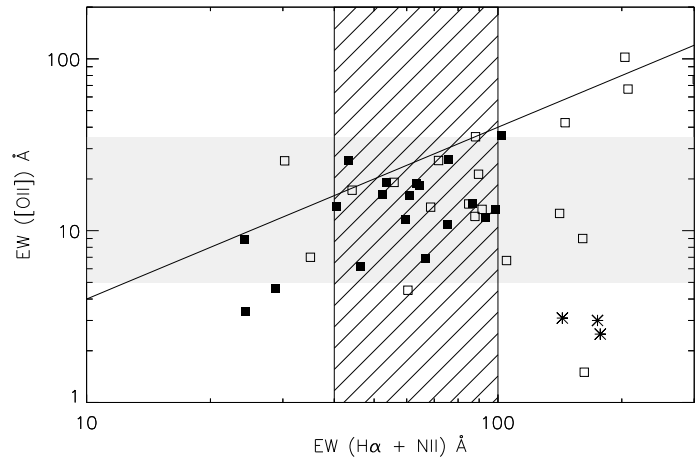}}
\caption{EW(OII) vs. EW(H$\alpha$+NII), filled squares e(a) galaxies,
open squares non--e(a) galaxies, stars Seyferts). The black line
corresponds to EW(OII)=0.4 EW(H$\alpha$+NII) found for nearby field
galaxies by \protect\cite{Kennicutt 1992}. The hatched vertical region shows the
location of VLT measurements of the \iso--HDF-S galaxies,
the shaded horizontal region the range from the \iso-CFRS sources. The 
intersection of the two bars
represents the inferred location  of the VLT \iso--HDF-S galaxies,
indistinguishable from the dusty, luminous, e(a) galaxies.  Figure taken from 
Rigopoulou et al. 2000}\label{fig:vlt}
\end{center}
\end{figure}

The first is from the ELAIS survey and is simply a summary of spectral classifications from the campaigns using 2dF and the ESO telescopes,  Gruppioni et al. (in prep.) and La Franca et al. (in prep.), presented by \cite{Oliver et al. 2000a} and shown in Table \ref{tab:class}.   The completeness and homogeneity of this list is not taken into account and so this table only gives a very crude picture. With that caveat in mind we do see that one third of these, relatively bright, mainly $15\mu$m sources have AGN of one sort or another.  A second important result arises from the VLT spectroscopy of $15\mu$m \iso-HDF-South sources undertaken by \cite{Rigopoulou et al. 2000} and illustrated in Figure \ref{fig:vlt}. This indicates that many of the fainter 15$\mu$m sources, making up the peak in the differential counts are star-forming galaxies, hinting that the population mix may be changing as we move to fainter fluxes.

\begin{figure}
\begin{center}
{\includegraphics[width=8cm]{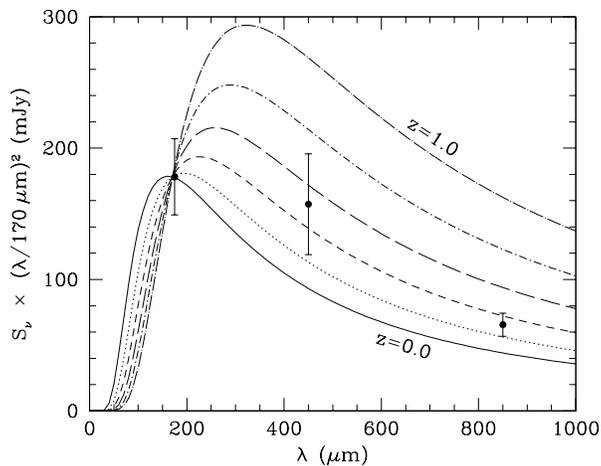}}
\caption{Average spectral energy distribution for a
FIRBACK sub-sample, from \protect\cite{Scott et al. 2000}.  The 170$\mu$m\ point is
the average of 10 FIRBACK sources, with error-bar
being the standard error from the scatter among them.
The 450$\mu$m\ and 850$\mu$m\ points
are from SCUBA observations.  For the sake of visual clarity 
$y$-axis has been multiplied by $x^2$.  The curves show emission from
modified blackbodies, normalised to the 170$\mu$m\ flux density, with
$T_{\rm d}\,{=}\,40\,$K and $\beta\,{=}\,1.5$, and for $z\,{=}\,0.0$
(solid line) up to $z\,{=}\,1.0$ in steps of 0.2.
Note that the shapes of these curves are degenerate in the combination
$(1+z)/T_{\rm d}$.}
\label{fig:sed}
\end{center}
\end{figure}

The third result comes from studies of the $175\mu$m sources.  These have been particularly hard to identify spectroscopically but the sub-mm photometric study of \cite{Scott et al. 2000} is instructive.  Figure \ref{fig:sed} shows the average fluxes of 10 sources compared with some simply SEDs.  From this it appears that the best fitting models would have the sources at relatively modest redshifts.

\section{CONCLUSIONS}

\subsection{What \iso\  source counts tell us}

\begin{itemize}
\item Strong significant evolution seen at all wavelengths observable to \iso.
\item Differential 15$\mu$m counts appear to steepen and then decline below about 1 mJy
\item Source counts particularly steep at 175$\mu$m
\item \iso\ reached the confusion limit at all wavelengths, easily at
the longest wavelengths
\item 10\% of the FIR background was resolved at long wavelengths
\item As much as 30\% of the FIR background may come from the sources 
detected at mid IR wavelengths.
\item Many models exist which might explain counts, though 175$\mu$m
counts are harder to explain than other bands
\end{itemize}

\subsection{What \iso\ survey follow-up tells us}

\begin{itemize}
\item Fainter mid IR sources appear to be predominantly star-forming,
rather than AGN
\item Brighter mid IR sources may contain have a higher fraction of AGN
\item Best current guess is that 175$\mu$m sources are $z<1$ although based on limited evidence
\end{itemize}

Forthcoming followup projects, notably X-ray surveys in \iso\ survey fields, should help
enormously to identify what populations make up these strongly evolving \iso\ sources.

\section*{ACKNOWLEDGMENT}

This paper is based on observations with {\em \iso}, an ESA project, with
instruments funded by ESA Member States (especially the PI countries:
France, Germany, the Netherlands and the United Kingdom) and with
participation of ISAS and NASA.
This work was in part supported by PPARC grant no.  GR/K98728
and  EC Network is FMRX-CT96-0068. 

I would like to thank the local organising committee for all the efforts
and for making this a very enjoyable meeting.

%%% 	REFERENCES (do not forget {}) %%%

\end{document}